# A Study of Rényi Entropy in pp Collisions by ALICE Data- A Comparison with UrQMD Simulation and AI-Assisted Modeling

**Swarnapratim Bhattacharyya**

**Department of Physics, New Alipore College, L Block, New Alipore, Kolkata 700053, India**

**Email: swarna_pratim@yahoo.com**

**Abstract**

In the present investigation a systematic study of order dependence and energy dependence of Rényi entropy calculated from the ALICE experimental data of charged-particle multiplicity distributions in pp collisions at the centre-of-mass energies $\sqrt{s}$ = 2.76, 7 and 13 TeV within the pseudo-rapidity interval $-0.8<\eta\leq0.8$ and transverse momentum range 0.15 GeV/c$<p_T<$10GeV/c has been carried out. ALICE collaboration compared the experimental multiplicity distribution of charged particles in pp collisions by PYTHIA 8.306 and EPOS-LHC model and hence Rényi entropy has also been evaluated from the corresponding PYTHIA 8.306 and EPOS-LHC model results reported by the ALICE Collaboration. An independent comparison has been carried out using UrQMD simulation generated for the same collision energies and kinematic acceptance by the author himself. In addition, an AI-assisted multiplicity distribution is considered as an auxiliary statistical reference for Rényi entropy estimation. In all the cases, the Rényi entropy exhibits a decreasing trend with increasing order q, indicating the presence of multifractality in pp collisions. The values of Rényi entropy for a particular order increases systematically with the increase of collision energy from $\sqrt{s}$ = 2.76TeV to 13 TeV, reflecting the increasing disorder and dynamical fluctuations of the particle-production process.

## Introduction

The charged-particle multiplicity distribution in proton–proton (pp) collisions at the Large Hadron Collider (LHC) is an important observable for investigating the multiparticle production mechanisms. It contains essential information about the statistical properties of particle production and the fluctuations occurring from one collision event to another. Charged-particle multiplicity distributions provide fundamental insight into the dynamics of particle production in high-energy collisions. Mathematically, the multiplicity distribution describes the probability of producing N number of charged particles in an individual collision event. This probability can be represented by the distribution function $P_N$, which satisfies the normalization condition $\sum_N P_N=1$. The shape and structure of the multiplicity distribution reflect the underlying fluctuations associated with various particle-production processes, including multiple parton interactions, color reconnection and parton fragmentation. Therefore, studying the fluctuation patterns encoded in the multiplicity distribution provides an opportunity to investigate the dynamics of particle production and the properties of strongly interacting matter in small collision systems.

Beyond its role in characterizing particle production, the multiplicity distribution provides valuable insight to investigate the entropy of the collision system. Entropy reflects the degree of disorder and the underlying randomness in the particle production processes. In the context of high-energy collisions, such as those involving hadrons or heavy ions, entropy serves as a fundamental thermodynamic observable that reflects the microscopic dynamics of the system. It is intimately connected to the particle multiplicity, phase-space distributions and the number of accessible microstates, thereby characterizing systems with a large number of degrees of freedom. In relativistic collisions, entropy is often associated with the early-stage dynamics and it provides valuable insight into the initial conditions, thermalization and possible formation of a strongly interacting medium such as the quark–gluon plasma. In recent years, detailed studies of entropy evolution in high-energy collisions have attracted significant attention among the physicists [1-5]. Various theoretical approaches have been developed to estimate the entropy associated with the final-state hadronic system. Among these, the method introduced by Simak et al. [6] has been widely adopted in studies of particle production and final-state entropy. The approach provides a practical framework for evaluating the entropy carried by the produced hadrons and has consequently been applied in a number of related analyses. Its use allows the entropy of the final state particles to be quantified in a systematic manner and facilitates comparisons of entropy-related observables across different collision conditions and systems. According to them, the entropy S can be obtained directly from the multiplicity distribution $P_N$ as given below

$$\mathbf{S} = -\sum_{\mathbf{N}} \mathbf{P_N} \ln \mathbf{P_N} \qquad (1)$$

The above expression represents the mathematical form of the well-known Shannon entropy [6-7]. The concept of Shannon entropy has found wide applicability across various scientific domains [8-10]. In heavy-ion collisions, Cao and Hwa initially introduced Shannon entropy to evaluate chaotic phenomena arising from branching-induced particle production [11]. Subsequently, Y. G. Ma adapted this approach to propose an innovative indicator for the nuclear liquid–gas phase transition [12]. The relation between Shannon entropy and the equation of state of nuclear matter originates from the thermodynamic nature of particle multiplicity fluctuations. These fluctuations, governed by the equation of state, determine the form of the multiplicity distribution, and hence the Shannon entropy constructed from it reflects the underlying thermodynamic behavior of the system. As pointed out by Deng and Ma [1], Shannon entropy can serve as an alternative observable of the QGP phase transition because it is sensitive to changes in multiplicity fluctuations and correlations. Near the phase transition, the modification of particle production dynamics and the increase in effective degrees of freedom lead to significant changes in the multiplicity distribution, which are reflected in the behavior of Shannon entropy. In this context, it may be noted that Shannon entropy is widely used as an observable to investigate the microscopic complexity and disorder of a system. However, to obtain a deeper understanding of the fluctuations present in the particle production processes, the study of Rényi entropy has also attracted considerable attention. Rényi entropy provides additional insight into the characteristics of the multiplicity distribution of the produced particles. Białas and Czyż [13–15] investigated the applicability of Rényi entropy in thermodynamic systems, considering it as an extension of the conventional Shannon entropy. Defined by an order parameter q, Rényi entropy constitutes a generalized entropic measure that tends to Shannon entropy when q approaches one, as discussed in [13-15]. In its general form, Rényi entropy of order q is expressed as:

$$\mathbf{H_q}=\frac{\mathbf{1}}{\mathbf{1-q}}\mathbf{ln C_q} \qquad \text{………………………(2)}$$

The quantity $\boldsymbol{C_q}$ appearing above can be written in terms of the multiplicity distribution $P_N$ as:

$$\mathbf{C_q} = \sum_{\mathbf{N}}(\mathbf{P_N})^{\mathbf{q}} \text{……………………..(3)}$$

A key difference between Rényi entropy and conventional Shannon entropy is that Rényi entropy is constructed from moments of the probability distribution itself and not from moments of kinematic quantities or of multiplicity of the collision process. So while Shannon entropy, requires full knowledge of the multiplicity distribution, Rényi entropy for a given order q requires only the sum of probabilities raised to the q-th power, a quantity that can frequently be inferred directly from experimental observations. This ease of experimental determination has contributed to its wide adoption in many areas of physics [16-23] and makes it operationally simpler to obtain than Shannon entropy [24]. As highlighted by Ozawa and Javerzat [24], Rényi entropy is a single-parameter extension of Shannon entropy originally formulated in information theory [25]. Studying its variation with the order parameter allows extraction of generalized fractal dimensions, which can be used to investigate whether multiparticle creation displays multifractal features [26-27]. Hence, Rényi entropy analysis provides an efficient means to probe the dynamics of underlying multiparticle production mechanisms.

## Goal of the Study

The ALICE collaboration has measured the multiplicity distributions of charged particles for inelastic (INEL>0) proton-proton (pp) collisions in the central pseudo-rapidity region $|\eta| < 0.8$ and transverse momentum interval $p_T$ = 0.15–10 GeV/c, at $\sqrt{s}$ = 2.76, 7 and 13 TeV [28]. They have compared their experimental data with PYTHIA 8.306 and EPOS-LHC model.

The present goal of this study is to calculate the values of Rényi entropy of order 2 to 5 from the experimental multiplicity distribution of charged particles produced in pp collisions as presented in [28] and also from the multiplicity distribution available in the literature for PYTHIA 8.306 and EPOS model in pp collisions at centre-of-mass energies $\sqrt{s}$ = 2.76, 7 and 13 TeV within the pseudo-rapidity interval $-0.8<\eta\leq0.8$ and transverse momentum range 0.15 GeV/c $<p_T<$10 GeV/c [28]. The author has simulated pp collisions data at the same kinematic range as studied by ALICE collaboration by UrQMD model. Rényi entropy of different orders will also be calculated from the multiplicity distribution of charged particles simulated by UrQMD model at $\sqrt{s}$ = 2.76, 7 and 13 TeV. In addition, an AI-assisted multiplicity distribution has been considered as an auxiliary statistical reference for entropy estimation for the first time. The comparative study of ALICE data, PYTHIA 8.306, EPOS-LHC, UrQMD and the AI-assisted multiplicity distribution provides an information-theoretic characterization of multiplicity fluctuations and offers an additional quantitative means of assessing how different descriptions reproduce the statistical structure of charged-particle production in pp collisions at $\sqrt{s}$ = 2.76, 7 and 13 TeV .

This AI based approach is the first ever applied in this context offers a computationally efficient alternative to event-by-event analysis of pp collision data while maintaining a clear separation between three distinct categories: (i) experimentally measured data from ALICE, (ii) predictions from PYTHIA 8.306 and EPOS-LHC as analysed by ALICE collaboration (iii) Comparison with the results obtained from the analysis of UrQMD simulated events by the author himself. Maintaining this distinction is essential for transparent interpretation of the results.

## Data Analysis and Results

ALICE collaboration studied [28] the multiplicity distribution of charged particles in proton–proton (pp) collisions at $\sqrt{s}$ = 2.76, 7 and 13 TeV within the pseudo-rapidity interval $-0.8<\eta\leq0.8$ and transverse momentum range 0.15 GeV/c $<p_T<$ 10 GeV/c. To ensure that the data sample used in their analysis consisted solely of genuine inelastic pp collisions, a multiplicity cut was applied to the minimum-bias (MB) events, requiring each event to contain at least one charged particle within that specified kinematic range. This criterion effectively eliminates the single-diffractive events, which typically produce few particles confined to the forward rapidity region. For the selected event sample, the average charged-particle multiplicities measured by the ALICE collaboration are presented in Table 1. ALICE collaboration simulated pp collision data by PYTHIA 8.306 and EPOS model. They have used the PYTHIA 8.306 event generator with Monash-2013 tune to simulate pp collisions at $\sqrt{s}$ = 2.76, 7 and 13 TeV. PYTHIA 8.306 is tuned with Monash 2013 because its default parameters do not reproduce the measured charged-particle multiplicity distribution at LHC energies. The Monash tune adjusts the MPI and hadronization parameters to match the experimental multiplicity distribution and its energy dependence.

From their study the multiplicity distribution of charged particles in pp collisions at $\sqrt{s}$ = 2.76, 7 and 13 TeV has also been extracted and calculated values of average multiplicity for both the models are presented in table 1. Now to explore the multiparticle production mechanism in more details in the present paper, The author has calculated the values of Rényi entropy from the multiplicity distributions of charged particles produced in pp collisions at $\sqrt{s}$ = 2.76, 7 and 13 TeV [28] and tabulated the values in Table 2. Rényi entropy values have also been calculated from the multiplicity distribution of charged particles generated by PYTHIA 8.306 and EPOS-LHC model as studied by ALICE collaboration in the same publication [28]. Calculated values of Rényi entropy has been tabulated in table 2. From the table it may be noted that the PYTHIA 8.306 simulated values of Rényi entropy are lower than the corresponding experimental values whereas EPOs-LHC simulated values of Rényi entropy are higher than the corresponding ALICE experimental values. This indicates that the multiplicity distribution generated by PYTHIA 8.306 is narrower and less fluctuating than observed in data, whereas EPOS-LHC produces a broader distribution with larger fluctuations due to its core-corona and hydrodynamic components.

To investigate the mechanism of charged-particle production and the associated dynamical features of proton–proton collisions at the LHC energies in terms of Rényi entropy in more detail, a separate simulation study is required to provide a theoretical framework for interpreting the experimentally observed multiplicity distributions. In the present work, the experimental data of the ALICE Collaboration at the centre-of-mass energies $\sqrt{s}$ = 2.76, 7 and 13 TeV have been simulated using the Ultra-relativistic Quantum Molecular Dynamics model (version 3.3p1). The UrQMD model is a microscopic transport model based on the covariant propagation of hadrons, which describes multiple particle production through string excitation and fragmentation at high energies [29-31]. In terms of partonic interactions and string formation by Lund model [32], the early stages of collisions are described in UrQMD. The later stages of the collision have been described by the hadronic transport component of this model. Excitation and fragmentation of the string along with the decay of hadronic resonances are responsible for particle creation by UrQMD model. Hard collisions involving large momentum transfer have been implemented in this model through the PYTHIA program. For details about this model, readers are requested to consult [29-31]. Here it may be mentioned that the UrQMD model has been widely used by the physicists in high-energy physics as an event generator. The UrQMD model effectively reproduces particle production characteristics across a broad energy spectrum, ranging from the Bevalac accelerator domain to experiments conducted at RHIC and the LHC, in both heavy-ion and proton–proton (pp) collisions.

To investigate the charged-particle multiplicity distributions in proton–proton (pp) collisions within the framework of the UrQMD model, a large ensemble of minimum-bias (MB) events was generated at the centre-of-mass energies $\sqrt{s}$ = 2.76, 7 and 13 TeV [33]. The analysis of the UrQMD-generated events was performed within the pseudo-rapidity interval $-0.8 < \eta \leq 0.8$ and the transverse momentum range 0.15 GeV/c $<p_T<$ 10 GeV/c which is consistent with the kinematic coverage of the ALICE measurements [28]. To ensure a meaningful comparison with the experimental results, the same event-selection criteria was applied to the simulated sample by retaining only those minimum-bias events which contain at least one charged particle within the same kinematic acceptance as the experimental data. Consequently, the simulated event sample closely matches the characteristics of the experimental dataset, thereby minimizing potential biases associated with the differences in event selection. The average multiplicity of charged particles has now been calculated for the UrQMD simulated events at $\sqrt{s}$ = 2.76, 7 and 13 TeV and presented in table 1 along with the corresponding experimental values. Rényi entropy values have been calculated from the probability distribution of charged particles generated by UrQMD model at $\sqrt{s}$ = 2.76, 7 and 13 TeV and have been tabulated in table 2. Table 2 reflects that the UrQMD model yields higher Rényi entropy values than the ALICE experimental data.

For the first time in the present study, an AI-assisted procedure is employed solely to construct a synthetic charged-particle multiplicity distribution under the same kinematic conditions as those used in the ALICE measurements. The AI is not used as a microscopic event generator and does not model the underlying collision dynamics, particle-production mechanisms, or transport processes. Instead, its role is limited to generating a mathematically well-defined discrete probability distribution, $P_N$, whose multiplicity characteristics are constrained by the experimentally reported ALICE reference data. The resulting $P_N$ should therefore be understood as a synthetic statistical representation of the observed multiplicity behaviour rather than as a collection of simulated proton–proton collision events. The use of an explicit discrete probability distribution provides a direct and well-defined input for the information-theoretic analysis. Once the multiplicity distribution of charged particles has been evaluated, the Rényi entropy can be computed for different orders q directly from the corresponding probabilities and it possible to investigate the dependence of the Rényi entropy on the order q.

The AI-assisted distribution is constructed under the reference kinematic conditions corresponding to charged particles in the mid-rapidity region, $\eta| < 0.8$, $p_T = 0.15 – 10$ GeV/c and event selection criteria of ALICE experimental data was also considered here. The resulting multiplicity distribution is a parametrically defined synthetic distribution constrained by the experimentally reported multiplicity characteristics and is not obtained from a transport model or conventional event-generator calculation. Consequently, the Rényi entropies derived from this distribution quantify the statistical structure of the constructed multiplicity probabilities, rather than providing predictions of the microscopic dynamics of proton–proton collisions. In this sense, the AI-assisted procedure serves as a transparent and reproducible means of constructing the probabilistic input required for the present information-theoretic analysis, while maintaining a clear distinction between experimentally constrained statistical modelling and dynamical collision simulation.

This approach allows for a rigorous calculation of average multiplicity and Rényi entropy directly from discrete multiplicity distribution. The calculated values of average multiplicity for the AI assisted data have been given in table 1. Interestingly it may be pointed out that the AI assisted values of average multiplicity of charged particles is quite comparable to the experimental data at the total energy regime. The author has calculated the values of Rényi entropy from the multiplicity distribution generated by AI assistance and presented the values in table 2. Figure 1 -3 represent the variation of Rényi entropy $H_q$ with order q for the ALICE experimental data, PYTHIA 8.306 and EPOS simulation data by ALICE, collaboration, UrQMD simulation data by the author and AI assisted data for energy $\sqrt{s}$ = 2.76, 7 and 13 TeV in pp collisions. From table 2 it can be seen that no single model reproduces the Rényi-entropy values of ALICE experimental data most closely at all three collision energies. Instead, the degree of correspondence varies with collision energy, suggesting that the multiplicity characteristics represented by the different models evolve differently with increasing collision energy. The AI-assisted results remain relatively close to the experimental entropy values across the studied energy range, although their agreement is not uniformly closer than that of the conventional event-generator models. This comparison therefore provides a useful energy-dependent assessment of how the different multiplicity distributions represent the entropy characteristics observed in the ALICE data. A clear and systematic decrease of the Rényi entropy has been observed when the order q is increased for all collision energies and for all the considered descriptions. The observed dependence of Rényi entropy on its order parameter in the experimental data provides evidence for the multifractal nature of charged-particle production in proton–proton (pp) collisions at the LHC energies. A similar order dependence of Rényi entropy is also observed in the corresponding simulated analyses performed using the PYTHIA, EPOS-LHC, and UrQMD event generators. Furthermore, the AI-assisted analysis exhibits a consistent order-dependent behaviour of Rényi entropy. The agreement of these observations with the experimental data suggests that the multifractal characteristics of charged-particle production are nicely reflected by the conventional event-generator simulation and the AI-assisted analysis. The table also shows a systematic dependence of the Rényi entropy on collision energy. For the ALICE data, $H_q$ increases with increasing centre-of-mass energy at each fixed value of order q. The same qualitative energy dependence is visible in the UrQMD, PYTHIA 8.306, EPOS-LHC and AI-assisted distributions. Physically, the increase of Rényi entropy with collision energy indicates that the accessible charged-particle multiplicity distributions become statistically richer as more collision energy is available for particle production. The increase in Rényi entropy with collision energy reflects the increasing statistical disorder and fluctuations of the collision system, indicating a more complex and less ordered particle-production process at higher energies. In this sense, the Rényi entropy provides a compact measure of the increasing diversity of multiplicity states with collision energy. The model-to-data comparison further demonstrates that the different event-generation approaches describe the multiplicity probability structure to different degrees. UrQMD and EPOS-LHC follow the energy-dependent behaviour of the ALICE entropy relatively closely over the investigated range, whereas PYTHIA 8.306 generally gives lower entropy values. The AI-assisted multiplicity distributions reproduce the qualitative decrease of $H_q$ with q and the overall increase of entropy with collision energy, while their degree of agreement with ALICE varies with energy and entropy order. Thus, the AI-assisted result can be used as a statistical comparison of the entropy derived from the constructed multiplicity distribution, but it should not be interpreted as an independent first-principles event-generator prediction.

From the present analysis it may be noted that all the descriptions exhibit the same qualitative trends of Rényi entropy in terms of order dependence and energy dependence. Rényi entropy is a useful observable for comparing experimental , model generated and the AI-assisted multiplicity distributions in pp collisions at the LHC energy. Thus, comparing the Rényi entropy derived from ALICE data with that obtained from UrQMD, PYTHIA, EPOS3, and the AI-assisted multiplicity distributions provides a quantitative framework for investigating particle production mechanism.

## Conclusions

The present study has investigated the order and energy dependence of Rényi entropy calculated from the charged-particle multiplicity distributions in pp collisions at $\sqrt{s}$ = 2.76, 7 and 13 TeV using ALICE experimental data. In addition to the ALICE data, the corresponding PYTHIA 8.306 and EPOS-LHC results reported by the ALICE Collaboration, independently generated UrQMD simulation and an AI-assisted synthetic probability distribution have been considered as complementary references. The principal findings of the investigation are summarized below.

- The systematic decrease of Rényi entropy with increasing order q demonstrates a clear order dependence of the multiplicity fluctuations, providing evidence for the multifractal character of particle production in pp collisions for the experimental, model simulated and AI assisted multiplicity distribution.
- The study of Rényi entropy at $\sqrt{s}$ = 2.76, 7 and 13 TeV for all the cases and for each order shows that Rényi entropy increases with increasing the energy of collision. The increase in Rényi entropy with collision energy reflects the enhanced fluctuations of the charged-particle multiplicity distribution at higher energies.
- The comparison of Rényi entropy evaluated from ALICE data with the corresponding PYTHIA 8.306 and EPOS-LHC results, UrQMD simulation carried out in the present study and the AI-assisted synthetic probability distribution provides a common entropy-based framework for exploring different descriptions of the multiplicity distribution. The comparison shows how differences in the underlying probability distributions are reflected directly in the corresponding Rényi entropy values.

The present investigation reflects the usefulness of Rényi entropy as a complementary statistical observable for studying the charged-particle multiplicity distributions in high-energy pp collisions by ALICE data, PYTHIA, EPOS-LHC and UrQMD simulated data and also by an analysis of AI-assisted synthetic distribution. This work presents the first-ever analysis of this kind and provides a basis for extending the study to different collision energies, collision systems, and kinematic conditions, which may help to explore the evolution of fluctuations, statistical disorder, and complexity in high-energy collision processes.

**Table 1**

| Average Multiplicity of Charged Particles $\langle \mathbf{N} \rangle$ | | | | | |
|---|---|---|---|---|---|
| **Energy of pp collisions $\sqrt{s}$** | ALICE experimental data | PYTHIA 8.306 By ALICE collaboration | EPOS-LHC By ALICE collaboration | UrQMD Simulation By the Author himself | AI-Assisted (Anchored to ALICE experiment) |
| **2.76 TeV** | 7.18$\pm$0.24 | 6.76$\pm$0.11 | 8.50$\pm$0.12 | 8.49$\pm$0.15 | 7.66$\pm$0.12 |
| **7 TeV** | 8.86$\pm$0.12 | 9.22$\pm$0.17 | 11.66$\pm$0.15 | 11.01$\pm$0.18 | 9.61$\pm$0.14 |
| **13 TeV** | 10.31$\pm$0.09 | 9.72$\pm$0.21 | 12.15$\pm$0.13 | 12.37$\pm$0.21 | 11.39$\pm$0.17 |

**Table 1.** Average charged-particle multiplicity $\langle \mathbf{N} \rangle$ in pp collisions at $\sqrt{s}$ = 2.76, 7 and 13 TeV. The ALICE experimental values obtained from [28] are compared with the corresponding UrQMD model results, while the PYTHIA 8.306 and EPOS-LHC values are obtained from simulations performed by the ALICE Collaboration [28]. The AI-assisted results are obtained from the AI-assisted multiplicity distributions anchored to the ALICE data.

**Table 2**

| Rényi Entropy values of pp collisions at $\sqrt{s}$ =2.76TeV | | | | | |
|---|---|---|---|---|---|
| Order of Rényi Entropy | ALICE experimental | PYTHIA 8.306 | EPOS-LHC | UrQMD model | AI Assisted value |
| 2 | 2.65±0.02 | 2.40±0.04 | 2.69±0.02 | 2.81±0.04 | 2.67±0.02 |
| 3 | 2.52±0.03 | 2.28±0.03 | 2.58±0.04 | 2.70±0.02 | 2.60±0.01 |
| 4 | 2.47±0.01 | 2.21±0.02 | 2.52±0.01 | 2.65±0.02 | 2.56±0.01 |
| 5 | 2.40±0.02 | 2.16±0.02 | 2.47±0.02 | 2.60±0.01 | 2.53±0.02 |
| **Rényi Entropy values of pp collisions at $\sqrt{s}$ =7 TeV** | | | | | |
| Order of Rényi Entropy | ALICE experimental | PYTHIA 8.306 | EPOS-LHC | UrQMD model | AI Assisted value |
| 2 | 2.84±0.05 | 2.77±0.02 | 3.06±0.03 | 3.04±0.04 | 2.89±0.04 |
| 3 | 2.70±0.02 | 2.65±0.02 | 2.95±0.02 | 2.92±0.03 | 2.82±0.04 |
| 4 | 2.63±0.03 | 2.58±0.03 | 2.88±0.02 | 2.85±0.02 | 2.77±0.02 |
| 5 | 2.58±0.01 | 2.53±0.01 | 2.84±0.02 | 2.80±0.02 | 2.74±0.01 |
| **Rényi Entropy values of pp collisions at $\sqrt{s}$ =13 TeV** | | | | | |
| Order of Rényi Entropy | ALICE experimental | PYTHIA 8.306 | EPOS-LHC | UrQMD model | AI Assisted value |
| 2 | 2.98±0.06 | 2.81±0.06 | 3.09±0.05 | 3.14±0.08 | 3.05±0.03 |
| 3 | 2.84±0.05 | 2.67±0.04 | 2.96±0.03 | 3.00±0.05 | 2.98±0.03 |
| 4 | 2.75±0.03 | 2.59±0.03 | 2.88±0.04 | 2.93±0.03 | 2.93±0.02 |
| 5 | 2.69±0.02 | 2.53±0.02 | 2.83±0.01 | 2.87±0.03 | 2.90±0.01 |
| | | | | | |

**Table 2.** Rényi entropy of orders q=2–5 calculated from the charged-particle multiplicity distributions in pp collisions at $\sqrt{s}$ = 2.76, 7, and 13 TeV for ALICE experimental data [28], PYTHIA 8.306 simulated data by ALICE collaboration [28], EPOS-LHC simulated data by ALICE collaboration [28], UrQMD simulation in the present analysis, and AI-assisted multiplicity distributions. For all three collision energies, the Rényi entropy decreases with increasing order q, while its values generally increase with collision energy. The results also reveal differences among the experimental data, model predictions, and AI-assisted estimates.

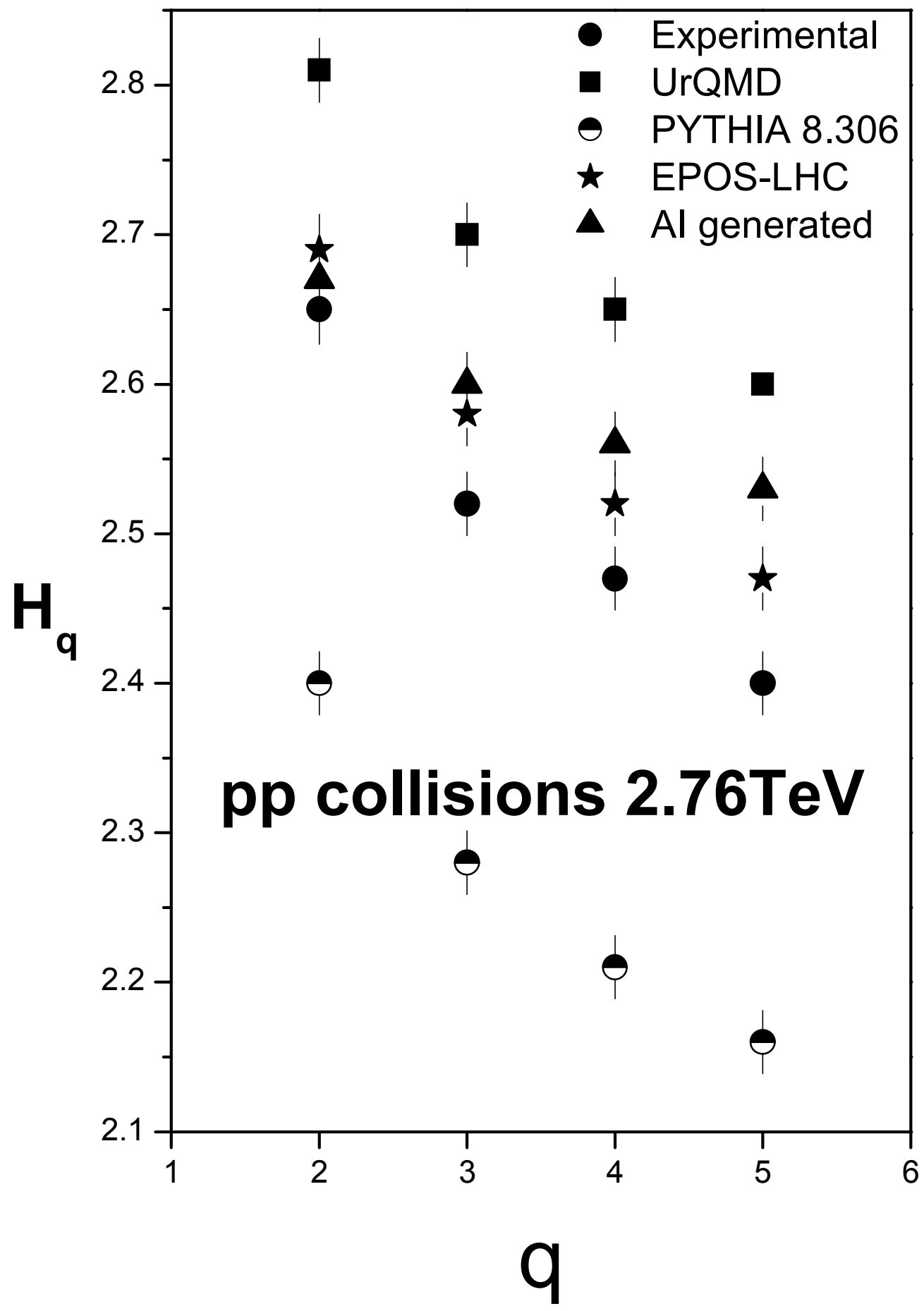

**Figure 1.** Rényi entropy $H_q$ as a function of the order q calculated from the charged-particle multiplicity distributions in pp collisions at $\sqrt{s}$ = 2.76 TeV, of ALICE experimental data, together with the corresponding simulation of PYTHIA, EPOS-LHC, UrQMD model and AI-assisted synthetic distributions. The ALICE experimental values obtained from [28] are compared with the corresponding UrQMD model results, while the PYTHIA 8.306 and EPOS-LHC values are obtained from simulations performed by the ALICE Collaboration [28]. The figure illustrates the systematic decrease of Rényi entropy with increasing order q for all the considered descriptions.

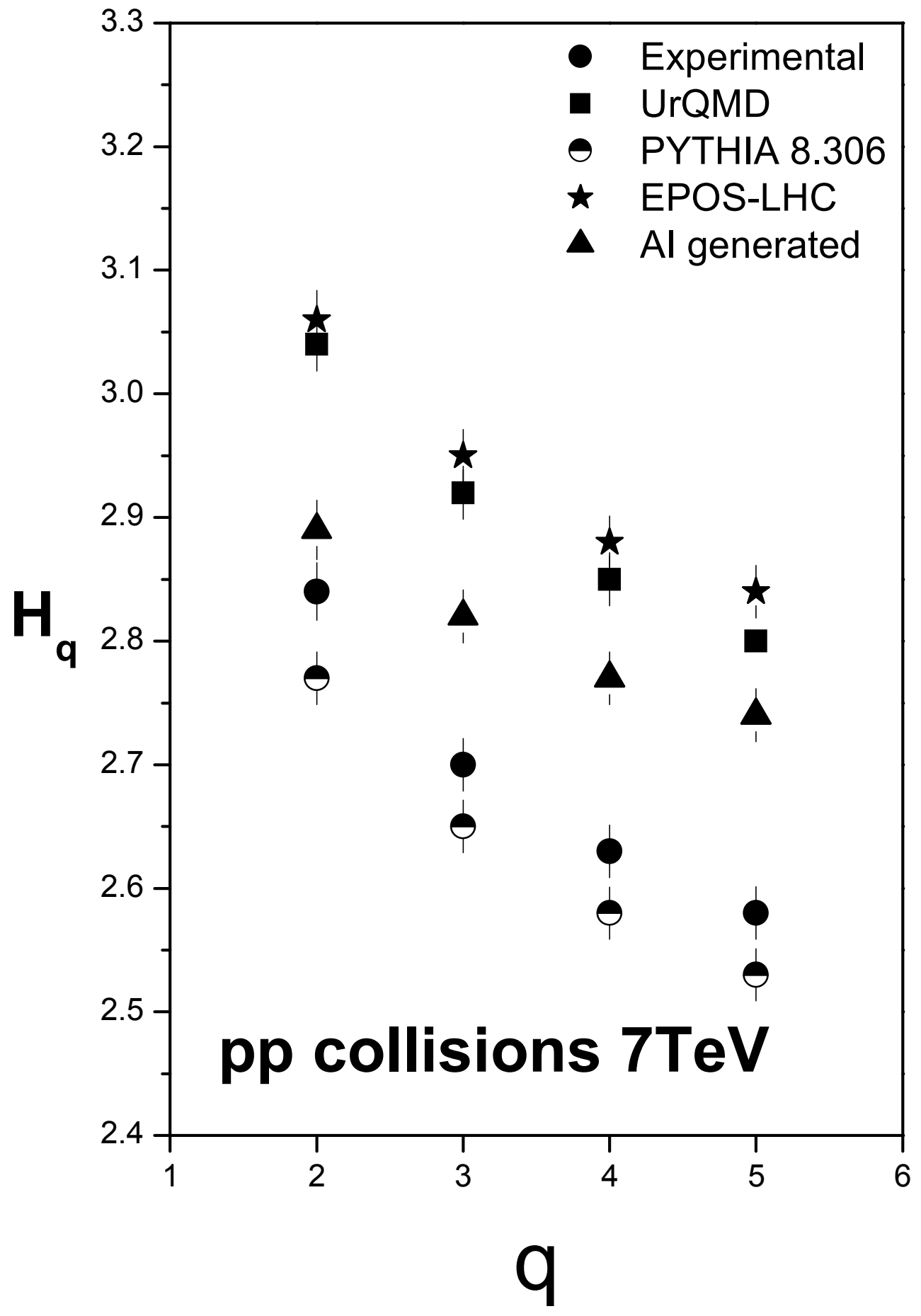

**Figure 2.** Rényi entropy $H_q$ as a function of the order q calculated from the charged-particle multiplicity distributions in pp collisions at $\sqrt{s}$ = 7 TeV, of ALICE experimental data, together with the corresponding simulation of PYTHIA, EPOS-LHC, UrQMD model and AI-assisted synthetic distributions. The ALICE experimental values obtained from [28] are compared with the corresponding UrQMD model results, while the PYTHIA 8.306 and EPOS-LHC values are obtained from simulations performed by the ALICE Collaboration [28]. The figure illustrates the systematic decrease of Rényi entropy with increasing order q for all the considered descriptions.

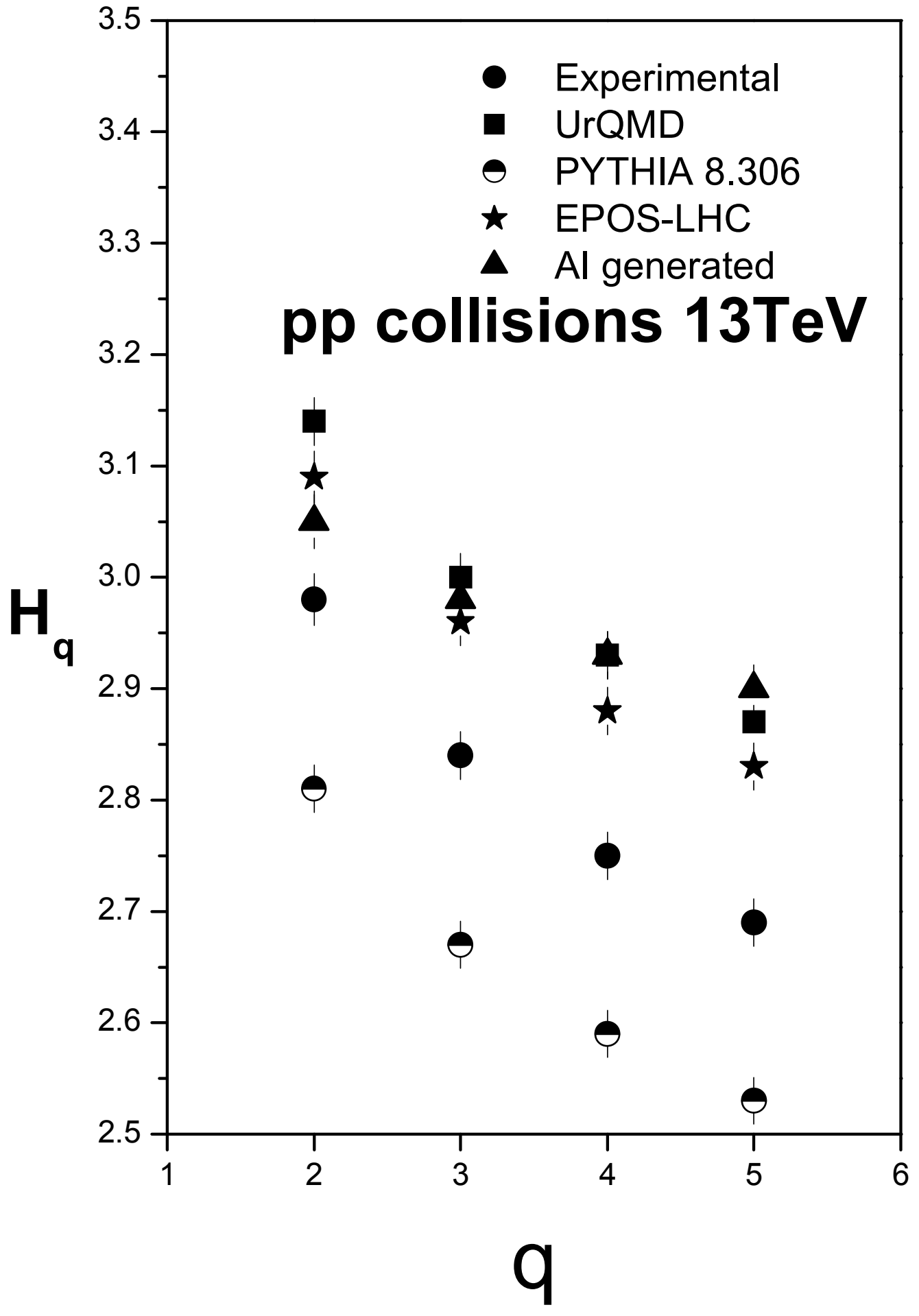

**Figure 3.** Rényi entropy $H_q$ as a function of the order q calculated from the charged-particle multiplicity distributions in pp collisions at $\sqrt{s}$ = 13 TeV, of ALICE experimental data, together with the corresponding simulation of PYTHIA, EPOS-LHC, UrQMD model and AI-assisted synthetic distributions. The ALICE experimental values obtained from [28] are compared with the corresponding UrQMD model results, while the PYTHIA 8.306 and EPOS-LHC values are obtained from simulations performed by the ALICE Collaboration [28]. The figure illustrates the systematic decrease of Rényi entropy with increasing order q for all the considered descriptions.

**Declaration of competing interest**

The author declares that he has no known competing financial interests or personal relationships that could have appeared to influence the work reported in this paper.

**Data availability**

In this paper, the simulated data have been generated using the UrQMD model simulation method. All data generated or analyzed during this study are included in this article.

The AI assisted multiplicity distributions were generated with Python code developed with assistance from Meta AI. The random seeds, CSV files $P_N$ vs N, and analysis scripts are available from authors on request. Scientific interpretation and conclusions are solely those of author.